\documentclass[12pt]{article}
\RequirePackage{lineno}
\usepackage{graphicx}
\usepackage{amsmath}
\usepackage{latexsym}
\usepackage{dcolumn}
\usepackage{bm}
\usepackage{float}
\usepackage{graphicx,here}
\usepackage{a4wide}
\usepackage{setspace}
\usepackage[numbers,sort&compress]{natbib}

\emergencystretch=\maxdimen
\begin{document}
\begin{spacing}{1.2}

\title{\bf Exfoliation of 2D van der Waals crystals in ultrahigh vacuum for interface engineering}

\author
{Zhenyu Sun$^{1,4}$, Xu Han$^{1,2}$, Zhihao Cai$^{1,4}$, Shaosheng Yue$^{1,4}$, \and Daiyu Geng$^{1,4}$, Dongke Rong$^{1,3}$, Lin Zhao$^{1,4}$, Yi-Qi Zhang$^{1,4}$, \and Peng Cheng$^{1,4}$, Lan Chen$^{1,4,5}$, Xingjiang Zhou$^{1,4}$, Yuan Huang$^{2\ast}$, \and Kehui Wu$^{1,4,5,6\ast}$, and Baojie Feng$^{1,4,6\ast}$\\
\\
\normalsize{$^{1}$Institute of Physics, Chinese Academy of Sciences, Beijing 100190, China}\\
\normalsize{$^{2}$Advanced Research Institute of Multidisciplinary Science,}\\
  \normalsize{Beijing Institute of Technology, Beijing 100081, China}\\
\normalsize{$^{3}$School of Science, China University of Geosciences, Beijing 100083, China}\\
\normalsize{$^{4}$School of Physical Sciences, University of Chinese Academy of Sciences,}\\
  \normalsize{Beijing 100049, China}\\
\normalsize{$^{5}$Songshan Lake Materials Laboratory, Dongguan, Guangdong 523808, China}\\
\normalsize{$^{6}$Interdisciplinary Institute of Light-Element Quantum Materials and Research Center}\\
  \normalsize{for Light-Element Advanced Materials, Peking University, Beijing, 100871, China}\\
  \normalsize{}\\
\\
\\
\normalsize{$^\ast$Corresponding author. E-mail: yhuang@bit.edu.cn;khwu@iphy.ac.cn}\\
  \normalsize{bjfeng@iphy.ac.cn}\\
}

\date{}

\maketitle

\clearpage

{\bf Two-dimensional (2D) materials and their heterostructures have been intensively studied in recent years due to their potential applications in electronic, optoelectronic, and spintronic devices. Nonetheless, the realization of 2D heterostructures with atomically flat and clean interfaces remains challenging, especially for air-sensitive materials, which hinders the in-depth investigation of interface-induced phenomena and the fabrication of high-quality devices. Here, we circumvented this challenge by exfoliating 2D materials in an ultrahigh vacuum. Remarkably, ultraflat and clean substrate surfaces can assist the exfoliation of 2D materials, regardless of the substrate and 2D material, thus providing a universal method for the preparation of heterostructures with ideal interfaces. In addition, we studied the properties of two prototypical systems that cannot be achieved previously, including the electronic structure of monolayer phospherene and optical responses of transition metal dichalcogenides on different metal substrates. Our work paves the way to engineer rich interface-induced phenomena, such as proximity effects and moir\'{e} superlattices.}
~\\
\maketitle


\section{Introduction}
Since the discovery of graphene in 2004, the study of two-dimensional (2D) materials and heterostructures has become one of the most active fields, as they hold great promise for next-generation quantum devices~\cite{LiuY2016,Novoselov2016}. To date, a wide variety of 2D materials have been realized, including transition metal dichalcogenides (TMDCs)~\cite{Manzeli2017}, hexagonal boron nitride~\cite{Roy2021}, and black phosphorus~\cite{XiaF2019}. Compared with bulk materials, 2D materials have a much higher surface-area-to-volume ratio. Therefore, the interfaces between 2D materials and substrates or other 2D materials can induce various exotic properties, which provides an effective method for manipulating the properties of 2D heterostructures through interface engineering. One example is proximity effects, including proximity-induced superconductivity~\cite{MorpurgoAF1999,WangM2012}, magnetism~\cite{Hellman2017}, and topology~\cite{Shoman2015}. When a topological insulator or ferromagnetic atomic chain is placed on a superconductor, the coexisting superconducting and topological orders can give rise to topological superconductivity, which hosts the long-sought Majorana fermions~\cite{Fu2008,Nadj-Perge2014}. Another example is interface-enhanced long-range order, such as superconductivity~\cite{WangQ2012}, magnetism~\cite{Hellman2017}, and charge density waves~\cite{Xi2015}. In addition, the moir\'{e} superlattices generated by stacking 2D materials with specific twist angles have attracted much attention in recent years. Periodic moir\'{e} potentials can lead to strong electron correlation effects, such as superconductivity~\cite{CaoY2018} and Wigner crystal states~\cite{ReganEC2020,Xu2020}.

The rapid development of 2D materials requires the preparation of high-quality ultrathin samples, which has spawned a variety of fabrication techniques ranging from bottom-up approaches such as molecular beam epitaxy (MBE) and chemical vapor deposition (CVD) to top-down approaches such as mechanical exfoliation. For bottom-up approaches, the growth parameters of each system need to be optimized, which is time-consuming and inefficient. In addition, only a limited number of substrates can be used to grow specific 2D materials, and only specific twist angles with the substrates can be obtained depending on the growth dynamics. These facts greatly limit the ability to manipulate the properties of 2D materials via interface engineering. Top-down approaches provide greater flexibility than bottom-up approaches. For layered van der Waals crystals, in principle, ultrathin flakes can be obtained on arbitrary substrates with controlled twist angles by liquid-phase or dry exfoliation.

However, traditional mechanical exfoliation is performed in air or glove boxes, and contaminants are inevitable compared to the ultrahigh vacuum (UHV) MBE technique. These contaminants are detrimental to the intrinsic properties of 2D materials, especially for air-sensitive materials, including black phosphorus~\cite{Favron2015}, transition metal tellurides~\cite{Ye2016}, and metal iodides~\cite{Shcherbakov2018}. This indicates the existence of large amounts of adsorbates, which questions the reliability of ex situ measurements of the intrinsic properties of 2D materials. The preparation of 2D materials and heterostructures with high-quality interfaces requires direct exfoliation of 2D materials in UHV to avoid possible contaminants. To this end, several methods have been developed. For example, the dry contact transfer method can be used to fabricate graphene nanoribbons and quantum dots \cite{RitterKA2008,RitterKA2009}; direct exfoliation of graphene has been realized on Si(111)-7$\times$7 reconstructed surfaces~\cite{Ochedowski2012} and epitaxial graphene~\cite{Imamura2020}. However, in-depth research on this technique, including the universality of other 2D materials and substrates, is still lacking.

Here, we demonstrate that vacuum exfoliation is a universal technique to obtain ultrathin 2D materials. The substrates for exfoliating 2D materials were obtained by standard surface preparation methods in UHV, such as annealing, ion sputtering, plasma treatment, and MBE growth. Assisted by ultraflat and clean surfaces, various 2D materials can be directly exfoliated by a simple ``stamp and peel-off'' procedure, forming an ideal interface with the substrates. The improved interface quality makes it possible to investigate their intrinsic properties by surface-sensitive techniques. In addition, this technique enables the investigation of interface-induced phenomena, such as the different optical responses of monolayer TMDCs on metal substrates. Our work not only enriches the diversity of realizable 2D heterostructures but also highlights the possibility of realizing rich exotic properties through interface engineering.

\section{Methods}
{\it 2.1. Preparation of the substrates}
\\
Single-crystal oxide substrates, including MgO, SrTiO$_3$, and Al$_2$O$_3$, were annealed at $>$1000 $^{\circ}$C for 6 hours in a tube furnace with appropriate oxygen flux. The as-prepared substrates show atomically flat surfaces, which were confirmed by atomic force microscope (AFM) measurements. The substrates were then transferred to the UHV exfoliation system ($\sim$10$^{-10}$ mbar), and annealed at $\sim$600 $^{\circ}$C to remove the surface adsorbates. The Si(111) substrates were annealed at $\sim$600$^{\circ}$C for 1 hour and then flashed to 1250 $^{\circ}$C several times to form the 7$\times$7 surface reconstruction. The cleanliness of the substrate surfaces was confirmed by in-situ low-energy electron diffraction (LEED) measurements. Polycrystalline metal films, including Au, Ag, Fe, and Cr, were prepared by growing high-purity metals onto the SiO$_2$/Si substrates in UHV. For Au and Ag films, Cr buffer layers are required to strengthen the adhesion with the substrates. Before growth, the SiO$_2$/Si substrates were annealed at 300 $^{\circ}$C to remove surface adsorbates. During growth, the substrate temperature was kept at room temperature. The thicknesses of the metal films were calibrated by a thickness monitor, and all metal films for UHV exfoliation were approximately 10-nm thick.\\

{\it\noindent 2.2. Mechanical exfoliation in UHV}
\\
Two vertically aligned sample stages were designed for the UHV exfoliation, as shown in Fig. 1(b). One stage is used to fix the substrates, and the other one is used to fix the freshly cleaved 2D van der Waals crystals. During the whole mechanical exfoliation process, the vacuum was kept at $\sim$10$^{-10}$ mbar. The exfoliation is typically conducted $\sim$10 min after the preparation of substrates.\\

{\it\noindent 2.3. In-situ characterization}
\\
After UHV mechanical exfoliation, the sample surface was checked by an optical microscope mounted outside the vacuum chamber. The surface quality was checked by in-situ LEED (OCI Microengineering Incorporation). The angle-resolved photoemission spectroscopy (ARPES) measurements were performed at room temperature with a SPECS PHOIBOS 150 analyzer and helium discharge lamp ($\sim$21.2 eV). Before ARPES measurements, the samples were transferred to the ARPES chambers without breaking the UHV. The energy resolution of ARPES was $\sim$15 meV.\\

{\it\noindent 2.4. Ex-situ characterization}
\\
AFM measurements were performed with a commercial facility (Oxford, Asylum Research Cypher S) in the tapping mode. X-ray photoelectron spectroscopy (XPS) measurements were performed using the Al K line (Thermo Scientific ESCALAB Xi+) with a base pressure $<$2$\times$10$^{-9}$ mbar. The XPS peaks were calibrated using the carbon 1s peak (284.8 eV). The Raman and photoluminescence (PL) measurements were carried out with an excitation wavelength of 532 nm (WITec alpha300R). The Raman peak of Si at 520.7 cm$^{-1}$ was used to calibrate each spectrum. The samples for optical characterizations were transferred from the UHV chamber to a glove box without exposure to air and then spin-coated with PMMA capping layers before taking out for optical characterizations. All AFM, XPS, Raman, and PL measurements were carried out at room temperature.

\section{Results and Discussion}
\subsection{Exfoliation of ultrathin 2D materials in UHV}

A schematic diagram of our method is illustrated in Fig. 1a-c. The system consists of three parts: preparation, exfoliation, and characterization. In the preparation chamber, ultraflat and clean substrates are prepared using standard surface preparation techniques. Freshly cleaved van der Waals crystals are then pressed onto the as-prepared substrates. The substrate temperature varies from room temperature to approximately 500 K, depending on the type of material. After several minutes, the crystals are quickly separated from the substrates, leaving ultrathin flakes on the substrates. The quality of the exfoliated samples is examined by an optical microscope through a viewport, as shown in Fig. 1c. The as-prepared ultrathin 2D materials can be transferred to other UHV measurement systems, such as LEED, ARPES, and scanning tunneling microscopy (STM), without breaking the vacuum.

\begin{figure*}[htb]
	\centering
	\includegraphics[width=16 cm]{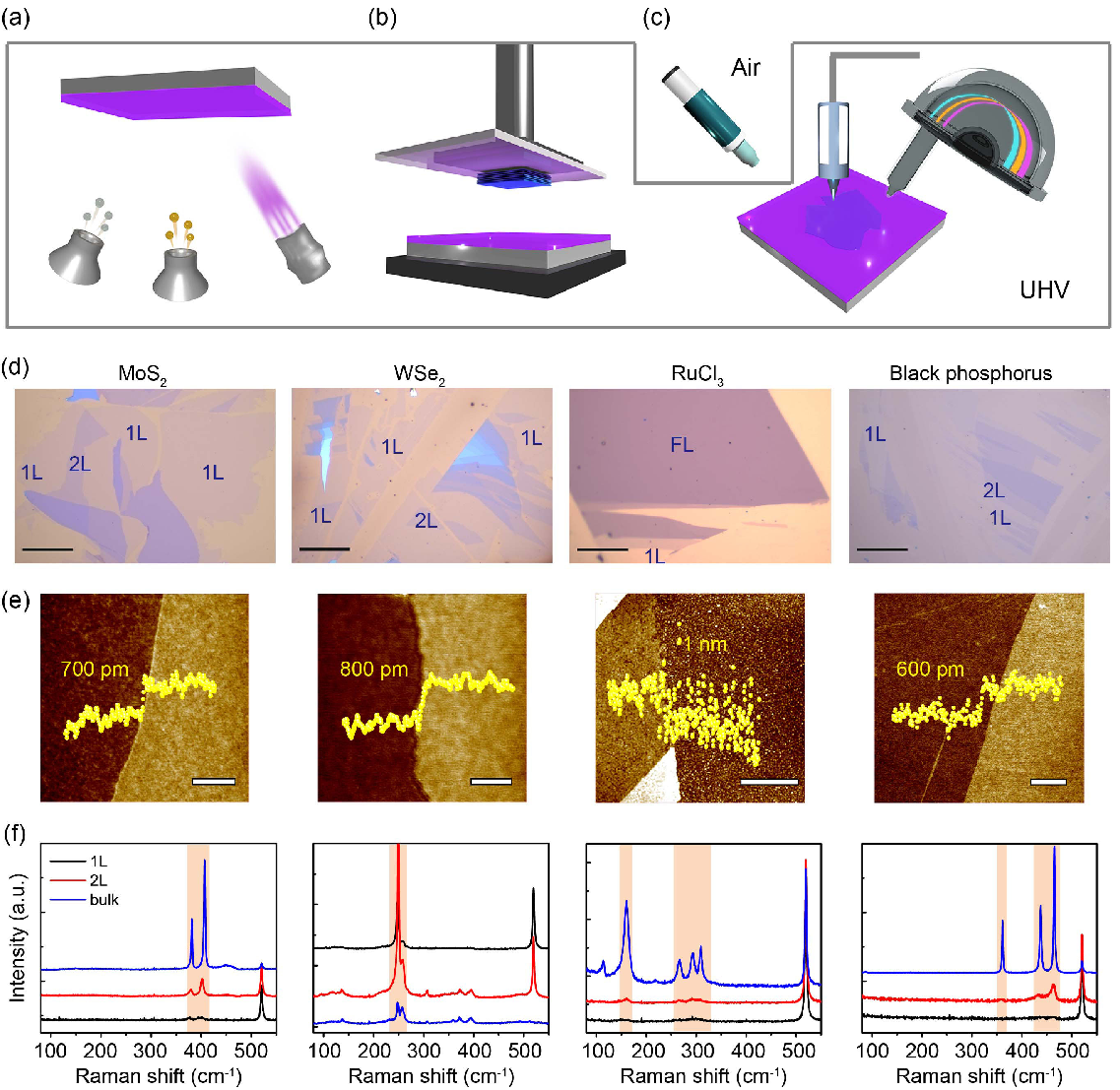}
	\caption{{\bf Mechanical exfoliation of 2D materials in UHV.} (a-c) Schematic drawing of the UHV exfoliation process, including preparation of ultraflat and clean substrate surfaces (a), mechanical exfoliation (b), and in situ characterization (c). (d) Optical images of large-scale 2D materials on polycrystalline Au obtained by the UHV mechanical exfoliation technique: MoS$_2$, WSe$_2$, RuCl$_3$, and black phosphorus. The 1L and 2L areas are indicated in each picture. Scale bar: 40 $\mu$m. (e,f) AFM and Raman spectroscopy characterization of the exfoliated 2D materials, respectively. Scale bar: 2 $\mu$m.}
\end{figure*}

First, we demonstrate that the Au-assisted exfoliation technique, which can be used to prepare large-scale monolayer 2D materials~\cite{Desai2016,Velicky2018,Huang2020,Liu2020}, is compatible with our system. Due to the strong adhesion of polycrystalline Au films, it is possible to exfoliate a variety of 2D materials at high yields. Figure 1d shows optical images of four typical 2D materials: MoS$_2$, WSe$_2$, RuCl$_3$, and black phosphorus. The large-scale monolayer and bilayer areas are marked in each picture. To further confirm the quality and thickness of the samples, ex situ AFM and Raman spectroscopy measurements were performed, and the results are shown in Fig. 1e,f. The surface is predominantly covered by ultrathin flakes, similar to previous works in air. However, distinct from our UHV exfoliation (see Fig. S1), the Au substrate degrades quickly in air, and the exfoliation yield declines rapidly after a few minutes of exposure to air~\cite{Velicky2018}.

Next, we show the ability to realize ultrathin 2D materials on single-crystal substrates with atomically flat interfaces. Such heterostructures are difficult to realize using traditional mechanical exfoliation techniques for the following reasons. First, for most materials, atomically clean and flat surfaces are only stable in ultrahigh vacuum. For example, the Si(111)-7$\times$7 surface is prone to adsorb contaminants when the vacuum is worse than $\sim$1$\times$10$^{-9}$ mbar. Second, even for stable materials, a large number of adsorbates or bubbles will be trapped at the interfaces during the exfoliation process, which may seriously affect their intrinsic properties~\cite{Gass2008,Purdie2018}. Third, the degradation of single-crystal surfaces, including the adsorption of contaminants or oxidization, tends to reduce the adhesion energy, which significantly reduces the exfoliation yield of ultrathin 2D materials~\cite{Huang2015}. Therefore, the preparation of 2D materials on single-crystal substrates with high-quality interfaces requires all preparation processes to be carried out in UHV. Previously, MBE has been proven to be an efficient method for the preparation of high-quality heterostructures, as exemplified by Bi$_2$Se$_3$/NbSe$_2$~\cite{WangM2012} and FeSe/SrTiO$_3$~\cite{WangQ2012}. However, as mentioned above, MBE has significant limitations compared to mechanical exfoliation, which calls for the development of vacuum exfoliation techniques.

We selected several typical single-crystal substrates, including MgO(100), SrTiO$_3$(100), Al$_2$O$_3$(0001), and Si(111). After pretreatment by UHV annealing or flashing, these substrates show atomically flat surfaces, as confirmed by AFM, STM, and LEED measurements (Fig. 2a-d). Surprisingly, the exfoliation yield increases significantly due to the improved surface quality of the substrates. Taking MoS$_2$ as an example, monolayer flakes can be obtained on all of these substrates, as shown in Fig. 2e-h. In addition to MoS$_2$, other 2D materials can also be exfoliated on these substrates, including graphene, FeSe, and Bi-2212, as shown in Fig. 2i-l. The quality of these ultrathin 2D materials was confirmed by Raman and AFM characterizations (Fig. S2-S4). The successful exfoliation of different types of ultrathin 2D materials on different substrates indicates the universality of this method, regardless of the 2D material and substrate. The increased exfoliation yield might originate from the enhanced coulombic attraction between 2D materials and the substrates since atomically flat and clean substrates increase the contact areas with 2D materials.

\subsection{Characterization of monolayer phosphorene}

Due to the improved quality of the surface and interface, the UHV exfoliation technique enables the study of exfoliated 2D materials by surface-sensitive techniques. Here, we take monolayer phosphorene as an example. Although few-layer phosphorene can be easily isolated from bulk black phosphorus, large-scale monolayer phosphorene is still difficult to realize. In addition, phosphorene is easy to degrade in air compared to UHV (see Fig. S5-S7), and exposing phosphorene to air or low-vacuum conditions will permanently degrade the surface quality. Such serious degradation hinders an in-depth investigation of its physical properties by surface-sensitive techniques. To date, only a few works have been reported on the optical and transport properties of monolayer phosphorene~\cite{Li2014,Li2017}, while direct studies of its structure and electronic properties by LEED and ARPES are still lacking.

Based on our UHV exfoliation technique, we isolated millimeter-sized monolayer phosphorene, as shown in the optical image in Fig. 3a; such large-scale samples enable measurements using conventional LEED and ARPES techniques. The thickness of the exfoliated phosphorene was confirmed by AFM and Raman spectroscopy characterizations, as shown in Fig. S8. The LEED pattern of the exfoliated monolayer phosphorene is displayed in Fig. 3c, which shows a rectangular structure. The absence of the (10) diffraction spots is due to the phase cancellation caused by the surface glide mirror symmetry~\cite{DaiZ2017}. The estimated lattice constants agree with those of bulk black phosphorus, indicating the absence of structural transition or charge density waves. The ARPES intensity map along the $\Gamma$-Y direction is shown in Fig. 3f. A hole-like parabolic band centered at the $\Gamma$ point is observed, which is consistent with previous calculations~\cite{Rudenko2015}. The valence band top is 0.2 eV below the Fermi level, slightly deeper than that of bulk black phosphorus. The downward shift of the hole-like bands might originate from electron doping effects of the Au substrate.

\subsection{Optical responses of monolayer TMDCs on different substrates}

Having confirmed the high quality of the exfoliated 2D materials by LEED and ARPES, we move on to show that the improved quality provides opportunities to realize rich interface-induced phenomena. Here, we study the optical responses of monolayer TMDCs on different metal substrates, including Au, Ag, Cr, and Fe. When TMDCs, such as MoS$_2$, come into contact with Au, the photoluminescence (PL) signal will be completely quenched because of the strong charge transfer at the interface~\cite{Velicky2018,Bhanu2014}, as shown in Fig. 4a. However, similar PL quenching effects have not been verified when monolayer TMDCs are placed on other metal substrates. The difficulty of such experiments lies in the air sensitivity of other metals. Unlike Au, which is relatively stable in air, the surfaces of Ag, Cr, and Fe will rapidly oxidize and adsorb contaminants under low-vacuum conditions. Therefore, using traditional mechanical exfoliation techniques, the exfoliated 2D material is in direct contact with metal oxides or contaminants rather than with the pristine metal surface. All of these difficulties can be avoided by the UHV exfoliation technique. Due to the improved surface quality of metal films under UHV, we successfully realized monolayer MoS$_2$ and WSe$_2$ on all of these substrates, as shown in Fig. S10.

The PL spectra of monolayer WSe$_2$ on different metal films are displayed in Fig. 4b. For WSe$_2$ on Au, the PL peak at $\sim$1.65 eV from the A exciton is completely quenched, consistent with previous reports~\cite{Velicky2018}. However, this peak was still visible when WSe$_2$ was placed on Ag, Cr, and Fe. In addition to the survival of the PL signal, we observed a significant redshift of the PL peak, which might be attributed to the enhanced trion photoluminescence~\cite{Mak2013}. Similar phenomena have also been observed in MoS$_2$, as shown in Fig. S11. There are three possible reasons for the survival of the PL signal of TMDCs on Ag, Cr, and Fe. First, these metals might have much weaker orbital hybridization with TMDCs than Au, and electrons in the excited states of TMDCs have a lower probability of transferring to substrates. Second, although Ag, Cr, and Fe hybridize with TMDCs, the transition of electrons from excited states to hybridized states is also radiative, leading to the survival of the PL signal. Third, since the PL spectra were acquired in ambient conditions, oxygen might penetrate TMDCs, oxidize the topmost metal layers, and decouple TMDCs from the substrates \cite{Velicky2020AMI}.

The Raman spectra of monolayer MoS$_2$ on different substrates are displayed in Fig. 4c. Similar to suspended MoS$_2$, the E$_{2g}^1$ and A$_{1g}$ peaks of MoS$_2$ on Fe and Cr are unimodal with a wavenumber spacing of $\sim$19 cm$^{-1}$~\cite{Li2012}. The shift of the Raman peaks is negligible, indicating a weak interaction of these substrates with MoS$_2$. When the substrate is switched to Au, the A$_{1g}$ peak is split, accompanied by a significant redshift and broadening of the E$_{2g}^1$ peak, in agreement with previous reports~\cite{Velicky2020}. The Raman behavior of MoS$_2$/Au might originate from charge transfer and local lattice deformation at the interface, as illustrated in Fig. 4d. In contrast with Au, the Ag substrate has different effects on the Raman behavior of MoS$_2$. The A$_{1g}$ peak does not split, while the E$_{2g}^1$ peak is split by $\sim$9 cm$^{-1}$. Since E$_{2g}^1$ is an in-plane vibrational mode, the splitting of this mode indicates a strong in-plane lattice distortion caused by the substrate. Similar results have been reported in strained~\cite{Lee2017} and twisted bilayer MoS$_2$~\cite{Quan2021}, but the splitting is smaller than the current case. In addition, all three Raman modes, especially 2LA(M), show significant enhancement on Ag. The enhancement of Raman peaks might originate from surface plasmon resonance, which has been reported in Ag nanoparticle-decorated MoS$_2$~\cite{Zhang2016}.

\section{Summary and outlook}

To summarize, we report a universal UHV exfoliation technique for realizing 2D materials and heterostructures with ideal interfaces. The improved surface quality grants access to study their intrinsic properties by surface-sensitive techniques, especially for air-sensitive materials. In addition, the universality of this technique provides unprecedented opportunities to realize rich interface-induced properties. Due to the different optical responses of TMDCs on metal substrates shown above, a number of intriguing properties can be readily expected. For example, the superconducting critical temperature of monolayer FeSe grown on SrTiO$_3$(001) is enhanced to $>$60 K compared with bulk FeSe. However, epitaxial FeSe can only grow along the crystallographic direction of SrTiO$_3$(001). With the UHV exfoliation technique, one can easily isolate ultrathin FeSe on SrTiO$_3$ or other substrates with controlled twist angles (see Fig. S14). In-depth investigations of these heterostructures may not only shed light on the mechanism for the enhanced superconductivity but also be conducive to realizing superconductors with higher critical temperatures.
~\\

{\bf\noindent Acknowledgements}
This work was supported by the Ministry of Science and Technology of China (Grant No. 2018YFE0202700), the National Natural Science Foundation of China (Grants No. 11974391, No. 11825405, No. 1192780039, and No. U2032204), the Beijing Natural Science Foundation (Grant No. Z180007), the International Partnership Program of Chinese Academy of Sciences (Grant No. 112111KYSB20200012), and the Strategic Priority Research Program of Chinese Academy of Sciences (Grants No. XDB33030100).
~\\

{\bf\noindent Author Contributions}
Baojie Feng, Kehui Wu, and Yuan Huang conceived the research. Baojie Feng, Kehui Wu, Yuan Huang, and Zhenyu Sun designed the UHV exfoliation system; Zhenyu Sun, Xu Han, Zhihao Cai, Shaosheng Yue, Daiyu Geng, Dongke Rong performed the characterization experiments; all authors contributed to the discussion of the data and writing of the manuscript.
~\\

{\bf\noindent Competing Financial Interests}
The authors declare no competing financial interests.
~\\

\begin{figure*}[htb]
	\centering
	\includegraphics[width=16 cm]{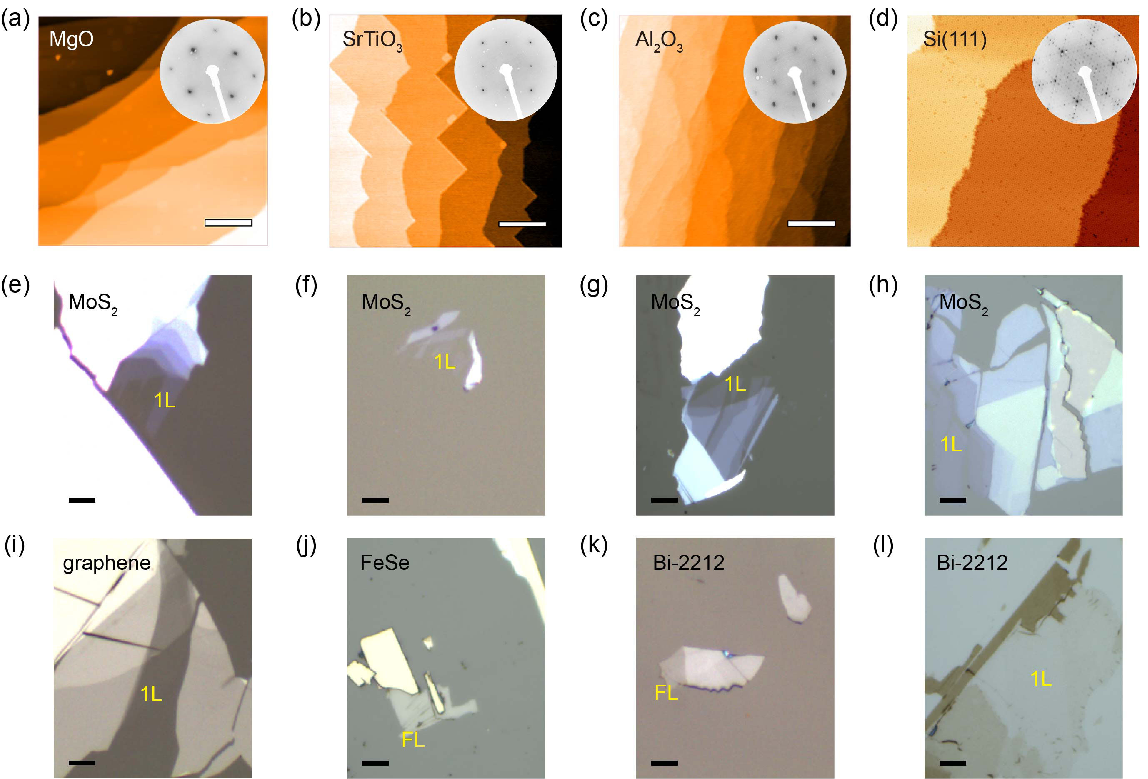}
	\caption{{\bf Exfoliation of ultrathin 2D materials on single-crystal substrates.} (a-c) AFM images of MgO(100), SrTiO$_3$(100), Al$_2$O$_3$(0001), respectively. (d) STM image of Si(111). The LEED pattern of each substrate is shown in the inset of (a-d). Scale bar: 400 nm. (e-h) Optical images of ultrathin MoS$_2$, graphene, Bi-2212, and FeSe. Scale bar: 5 $\mu$m.}
\end{figure*}

\begin{figure*}[htb]
	\centering
	\includegraphics[width=10 cm]{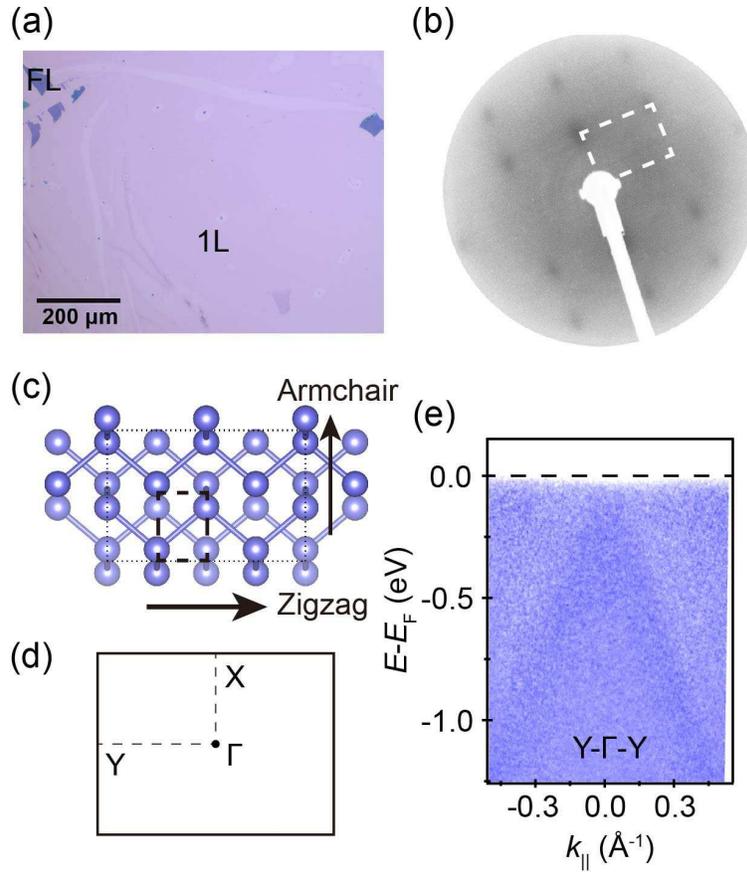}
	\caption{{\bf LEED and ARPES characterization of monolayer phosphorene.} (a) Optical image of UHV-exfoliated phosphorene on Au. The lateral size of monolayer phosphorene is on the order of millimeters. (b) LEED pattern of monolayer phosphorene exfoliated in UHV. (c,d) Crystal structure and Brillouin zone of monolayer phosphorene. (e) ARPES intensity plot of monolayer phosphorene along the $\Gamma$-Y direction measured at room temperature. The energy of the incident photon is 21.2 eV. The valence band maximum is located at the $\Gamma$ point.}
\end{figure*}

\begin{figure*}[htb]
	\centering
	\includegraphics[width=10 cm]{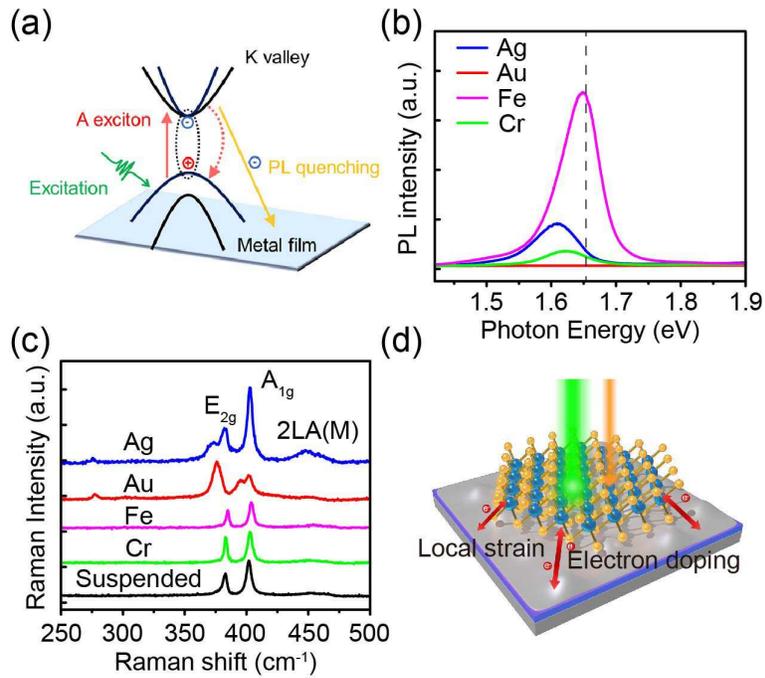}
	\caption{{\bf Optical responses of monolayer MoS$_2$ and WSe$_2$ on different metal films.} (a) Schematic drawing of the PL quenching process. The transition of electrons in the excited states of TMDCs to the substrate is nonradiative. (b) PL spectra of monolayer WSe$_2$ on Au, Ag, Fe, and Cr films. The vertical dashed line indicates the peak position of the suspended sample. (c) Raman spectra of monolayer MoS$_2$ on Au, Ag, Fe, and Cr. The suspended monolayer MoS$_2$ is shown for comparison. (d) Schematic drawing of monolayer TMDCs on metal substrates. Local strain and electron doping can affect the Raman behavior.}
\end{figure*}

\end{spacing}
\end{document}